# Integrated Multiport Back-to-Back Power Converter for Type-4 Wind Turbine Generator with Hybrid Energy Storage System

Bang Le-Huy Nguyen[†§], Thai-Thanh Nguyen[†], Van-Long Pham[‡], Tuyen Vu[†], Mayank Panwar[§], Rob Hovsapian[§],
[†]*ECE Department, Clarkson University*, Potsdam, NY, USA
[‡]*Power Electronics Laboratory, Yokohama National University,* JAPAN
[§]*National Renewable Energy Research Lab*, Golden, CO, USA
nguyenbl@clarkson.edu, pham-long-jk@ynu.ac.jp, tnguyen@clarkson.edu,
tvu@clarkson.edu, mayank.panwar@nrel.gov, rob.hovsapian@nrel.gov

*Abstract*— **This paper proposes a novel integrated multi-port bidirectional back-to-back power converter for a type-4 wind turbine that accommodates a battery and supercapacitor for energy storage. The circuit topology reduces 4 switches compared to the traditional configuration. Moreover, owing to the dual-buck structure embedded in the phase leg, the circuitry has no short-circuit path, therefore it withstands short-circuited events for a much longer time than the normal phase-leg and prevents the reverse current in turn-off recovery. The use of a hybrid energy storage system with battery and supercapacitor helps smooth out the power output under wind gusts and stabilizes the DC-link voltage under grid fault conditions. The case studies are carried out with a 1.5 MW wind turbine system. Simulation results are provided for the theoretical validation.**

## I. Introduction

Renewable energy resources, especially wind turbines, are replacing traditional synchronous generators in the new generation capacity in recent years [1]. This displacement rises more requirements for renewable generation systems; those are the islanding operation and black-start capability. Hence, the renewable systems can perform bottom-up restoration and supply critical loads [2]. For variable speed wind turbine generators, the doubly-fed induction generation (DFIG) [3] and permanent magnet synchronous generator (PMSG) [4] are employed, and so-called type-3 and type-4 wind turbines, respectively.

This paper focuses on the power converter configuration for a type-4 wind turbine with an energy storage system [5]. Conventionally, separated power converters are installed together as shown in Fig.1. The type-4 wind turbine system requires a machine-side AC/DC converter, and a grid-side DC/AC converter to convert the wind power and provide it to the grid [6]. Two bidirectional DC/DC converters are employed to accommodate the battery and super-capacitor [7]. The supercapacitor is responsible for smoothing the charged and discharged currents of the battery [8]. If the 2-level 3-phase topology is used for AC/DC and DC/AC power converters, the 2-level 2-switch leg is used for the bidirectional DC/DC converter. The wind turbine and energy storage system require totally 15 active semiconductor switches accompanied with their gate-driver circuits, auxiliary power supplies, and protection circuits.

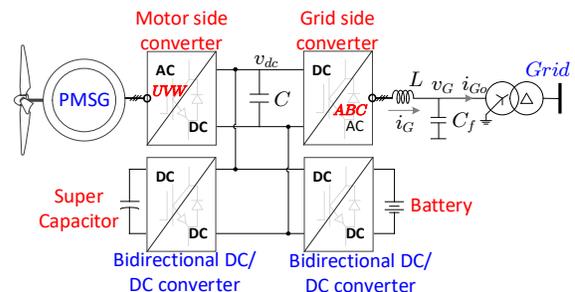

Fig. 1. Conventional configuration of type-4 wind turbine power electronics interfaces included energy storage system.

There are related efforts trying to derive a better power converter configuration for the wind turbine system. In [9], a reduced-switch power converter configuration with a multi-switch phase leg is proposed to integrate multiple PMSG wind turbines. Although the proposed configuration can reduce the number of active-switch, stacking more than 3 switches in the same phase leg significantly increases the voltage and current stresses on the active switches. In addition, the *n*-switch phase leg with *n* switching states requires (*n*-1) dead-time intervals. This practice considerably reduces the switching frequency of the power converter. The three-level neutral point clamped converter, modular multilevel converter, and H-bridge converter are configured as back-to-back converters in [10] for large wind turbines. The proposed hybrid converter structures are complicated and only suitable for multimegawatt wind turbines. Semi-active circuits such as VIENNA rectifier are employed in [11], however, these topologies are only suitable for low-power wind turbines. The three-port medium transformer configuration is proposed in [12] to accommodate the energy storage system. The main advantage of this configuration is to replace the traditional bulky low-frequency transformer. However, many power converters are employed to facilitate the operation of such a medium-frequency transformer. In [13], the reduced-switch power converter using a 6-switch active rectifier is proposed for two independent small-scale wind turbines. Since the connection from two wind turbines to the common power converter would be long practically. This configuration is unscalable.

This paper proposes a novel 9-switch back-to-back power converter connecting the PMSG and the grid. In addition, the 3-switch 2-port bidirectional DC/DC power converter is derived to connect the battery and supercapacitor to the DC-link of the back-to-back power converter as a hybrid energy storage system (HESS). Fig. 2 shows the proposed power converter configuration with four 3-switch 2-port phase legs. In each phase leg, two small inductors are embedded inside to prevent the short-circuited. Compared to [14], this structure reduces two inductors.

The contributions of the paper are listed in the following bullet points.
- The proposed power converter topology has less than 4 active switches compared to the conventional configuration of separated power converters.
- The proposed 3-switch phase-leg structure can withstand short-circuited at a much longer time compared to the normal 2-switch and 3-switch phase structure owing to the embedded inductors.
- The proposed topology enables the high-switching operation of active switches since the dead-time interval which prevents short-circuited can be minimized.
- The integration of supercapacitor and battery supports smoothing the output power.

The rest of the paper is organized as follows. In Section II, we analyze the operational principle of the proposed integrated multiport power converter. Section III describes the control strategy implemented for a wind turbine with an energy storage system. In Section IV, the simulation results are provided and discussed. We conclude the paper in Section V.

## II. OPERATIONAL PRINCIPLE OF PROPOSED INTEGRATED MULTIPORT POWER CONVERTER

The proposed power converter is derived from the bidirectional 3-switch leg [15], [16], where the active switch and diode are separated in a way similar to the dual-buck structure [17]. However, there is only an inductor embedded in between making two inductors in total for one 3-switch phase leg. The proposed leg also have 3 switching states [$S_1S_2S_3$] of **[110]**, **[101]** and **[011]**. Fig. 3 shows the shoot-through path of the proposed 3-

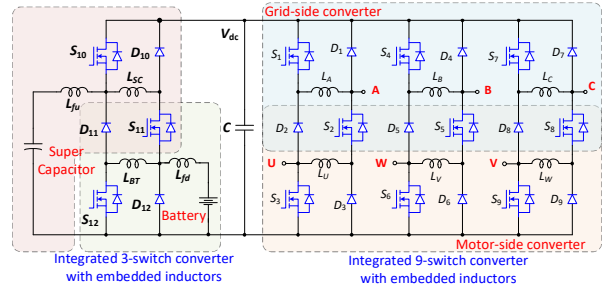
Fig. 2. Proposed integrated 9-switch and 3-switch power converter with embedded inductors for type-4 wind turbine included HESS.

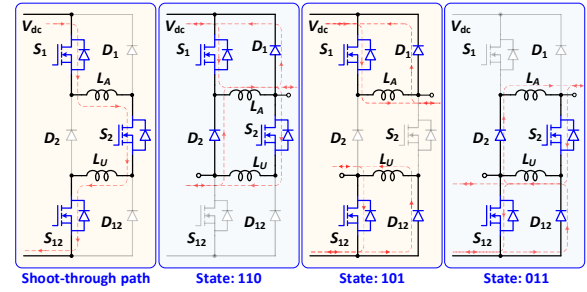
Fig. 3. Shoot-through path and switching states of the proposed three-switch leg with embedded inductors.

switch leg with their three-switching states.

### A. Shoot-Through Path.

When all three switches of a leg are accidentally turned on, the short-circuited on the DC-link occurs. The shoot-through path shown in Fig. 3 includes the DC-link, the switch $S_1$, the inductor $L_A$, the switch $S_2$, the inductor $L_U$, and the switch $S_3$. The short-circuit current is evaluated via the differential equation of an RL circuit as follows:

$$i_{sh}(t) = i_{L,0} e^{-\frac{R_{eq}}{L_{eq}}t} + \frac{V_{dc}}{R_{eq}}(1 - e^{-\frac{R_{eq}}{L_{eq}}t}) \quad (1)$$

where $i_{L,0}$ is the initial current flowing through the shoot-through path, $R_{eq}$ includes the resistance of turned-on switches, inductors, and the path, $L_{eq}$ includes two inductors ($L_A$, $L_U$), and the inductance of the shoot-through path. As can be seen, adding two inductors has limited the ramp rate of the shoot-through current. The dead-time inverter can be minimized to facilitate the high-frequency operation. The proposed 3-switch leg is more resilient compared to the normal phase leg.

The embedded inductors are designed to withstand the short-circuited when minimizing the dead-time intervals. Therefore, they should be very small at approximately 50 µH in this paper.

### B. Switching State: 110.

In this state, the upper and middle active switches ($S_1$, $S_2$) are turned on, whereas the lower active switch ($S_3$) is turned off. The diodes ($D_1$, $D_2$) freewheel if there are mismatches between the inductor currents and the output currents. The upper and lower outputs are connected to the positive pole of the DC-link.

## C. Switching State: 101.

In this state, the upper and lower active switches ($S_1$, $S_3$) are turned on, whereas the middle active switch ($S_2$) is turned off. The diodes ($D_1$, $D_3$) freewheel if there are mismatches between the inductor currents and the output currents. The upper and lower outputs are connected to the positive pole and negative pole of the DC-link, respectively.

## D. Switching State: 011.

In this state, the middle and lower active switches ($S_2$, $S_3$) are turned on, whereas the upper active switch ($S_1$) is turned off. The diodes ($D_2$, $D_3$) freewheel if there are mismatches between the inductor currents and the output currents. The middle and lower outputs are connected to the negative pole of the DC-link.

## E. Modulation Scheme.

The modulation scheme of the proposed 3-switch leg is depicted in Fig. 4. To avoid float-terminal, there is only one switch turned off at a time making 3 switching states as mentioned. The voltage reference of the upper output $r_u$ should be always larger than the voltage reference of the lower output $r_l$ as shown in Fig. 4. The switching signals of switches $S_1$ and $S_3$ are generated by comparison with the carrier wave, whereas the switching signal of $S_2$ is the exclusive disjunction (XOR) of $S_1$ and $S_3$ switching signals.

## F. Reference Signal Analysis.

The three-phase voltage references of the upper and lower outputs in the scale of [0; 1] can be expressed as follows:

$$\begin{cases} v_{u,a}(t) = 0.5 + 0.5m_u \sin(2\pi f_u t + \varphi_u) + v_{off,u} \\ v_{u,b}(t) = 0.5 + 0.5m_u \sin\left(2\pi f_u t + \varphi_u + \frac{2\pi}{3}\right) + v_{off,u} \\ v_{u,c}(t) = 0.5 + 0.5m_u \sin\left(2\pi f_u t + \varphi_u - \frac{2\pi}{3}\right) + v_{off,u} \end{cases}$$
(2)

$$\begin{cases} v_{l,a}(t) = 0.5 + 0.5m_l \sin(2\pi f_l t + \varphi_l) + v_{off,l} \\ v_{l,b}(t) = 0.5 + 0.5m_l \sin\left(2\pi f_l t + \varphi_l + \frac{2\pi}{3}\right) + v_{off,l} \\ v_{l,c}(t) = 0.5 + 0.5m_l \sin\left(2\pi f_l t + \varphi_l - \frac{2\pi}{3}\right) + v_{off,l} \end{cases}$$
(3)

where $v_{u,a}$, $v_{u,b}$, $v_{u,c}$ and $v_{l,a}$, $v_{l,b}$, $v_{l,c}$ are the three-phase voltage references for upper outputs (ABC) and lower outputs (UWV), respectively; $m_u$ and $m_l$ are the modulation indices; $f_u$ and $f_l$ are the frequency references; $\varphi_u$ and $\varphi_l$ are the initial phase references; and $v_{off,u}$ and $v_{off,l}$ are the three-phase offsets for upper outputs and lower outputs, respectively.

As aforementioned, the upper reference is always larger than or equal to the lower reference in the same leg.

$$v_{u,a} \geq v_{l,a}, v_{u,b} \geq v_{l,b}, v_{u,c} \geq v_{l,c} \quad (4)$$

To ensure this constraint, a positive value can be assigned for the upper offset $v_{off,u}$ and a negative value can be assigned for the lower offset $v_{off,l}$ so that the differences between upper and lower references can be increased by $(v_{off,u} - v_{off,l})$. In the scale of [0; 1], the upper and lower offsets are limited as follows.

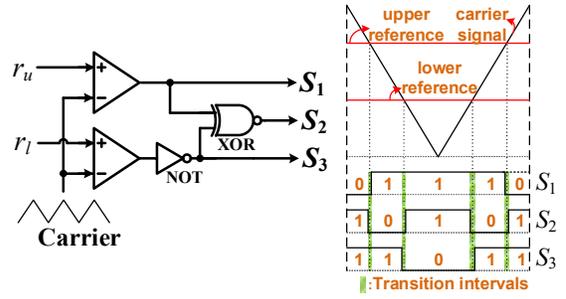

Fig. 4. Modulation scheme of the proposed 3-switch leg.

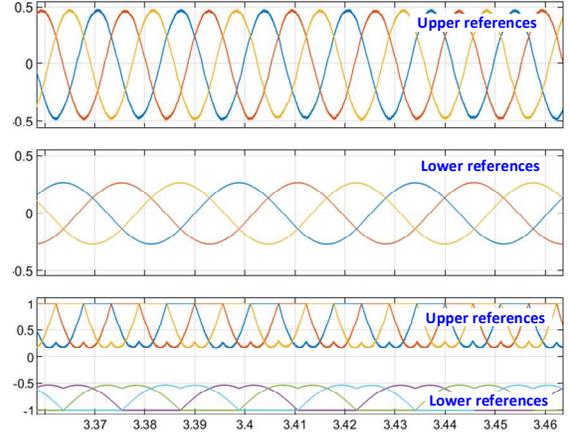

Fig. 5. The original upper references, the original lower references, and the upper and lower references with maximum and minimum offsets, respectively (from top-to-bottom).

$$1 - max_u \geq v_{off,u} \geq v_{off,l} \geq -min_l \quad (5)$$

where, $max_u = max(v_{u,a}, v_{u,b}, v_{u,c})$ and $min_l = min(v_{l,a}, v_{l,b}, v_{l,c})$. The value of $v_{off,u} = 1 - max_u$ and $v_{off,l} = -min_l$ are so-called the maximum and minimum offsets, respectively.

Fig. 5 compares the original waveforms of the upper and lower references without adding offsets and the upper and lower references with additional maximum and minimum offsets, respectively. As can be seen, the upper references are always larger than the lower references. Moreover, the maximum offset for upper references and minimum offset for lower references makes the discontinuous pulse width modulation signals (PWM), which can reduce the switching loss since there are no switching signals at 1 and 0 references.

Notably, the modulation indices of upper and lower references are limited to 0.5 to make sure that the upper references can be always larger than or equal to the lower references. As a result, the DC-link voltage should be doubled to maintain the output voltages are regulated at the same values as the conventional configurations. This is a trade-off of the proposed configuration.

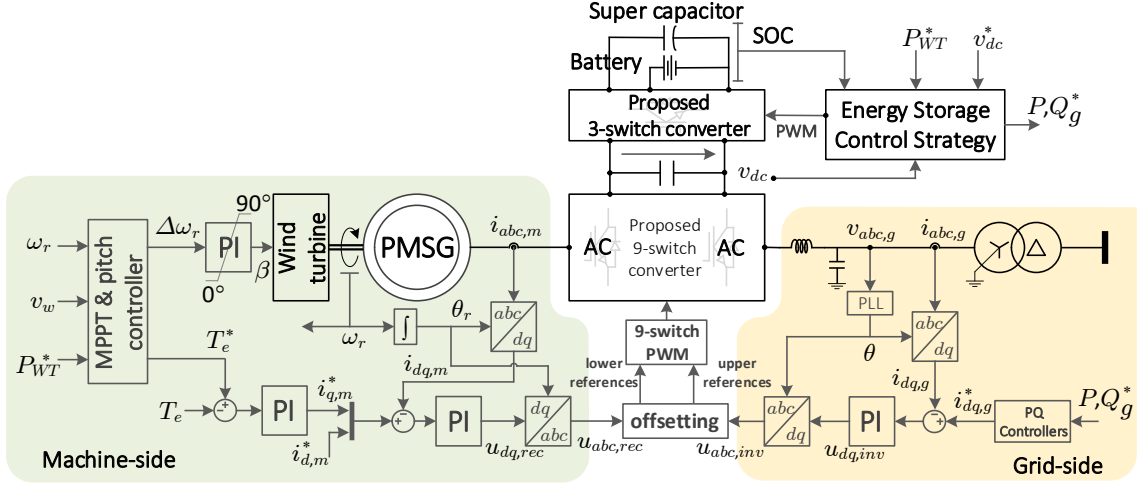

Fig. 6. Control strategy diagram of the PMSG wind turbine with integrated HESS using the proposed power converter structure.

## III. Control Strategy Implementation for Wind Turbine with Energy Storage System

The control strategy diagram of the PMSG wind turbine with integrated HESS using the proposed power converter structure is described in Fig. 6. Different from [18], where the grid-side and machine-side control loops are responsible for the torque and DC-link voltage regulation; herein, machine-side loops still control the electrical torque. However, the HESS is responsible for the regulation of DC-link voltage, while the grid-side control loops manage to dispatch the wind turbine system at the desired power.

The machine-side control loops regulate the wind turbine's electrical torque to track the maximum power point. The HESS is regulated to maintain the DC-link voltage at the desired level and to determine the output wind power dispatching to the grid. The grid-side control loops are to restrain the output power following the dispatch commands from the HESS control strategy. The outputs of machine-side and grid-side control loops are the output voltage references of the active rectifier and inverter, respectively. These voltage references are added to the appropriate offsets as expressed in Section II.E to generate the lower and upper references for the 9-switch PWM scheme.

The HESS control strategy is described in Fig. 7. The proposed control strategy includes the wind turbine dispatching scheme, the charge/discharge control strategy, and the state-of-change (SOC) balancing control. In this paper, we leave the wind turbine dispatching and the SOC balancing scheme for future works while only focusing on the charge and discharge mode control. Herein, the desired dispatch power is compared with the wind turbine power. Thereafter, the excessive power is charged into the battery and super-capacitor, whereas the lacking power is replenished by the battery and super-capacitor discharge. The battery only absorbs power following the constant-current and constant voltage modes, while the remainders are taken care of by the super-capacitor.

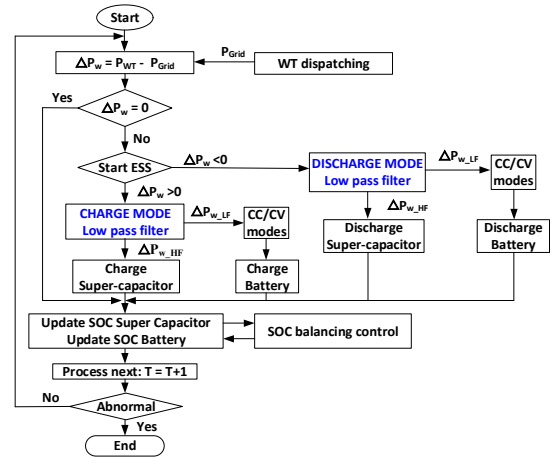

Fig. 7. Charge and discharge control strategy HESS.

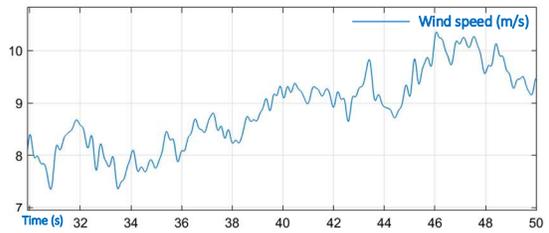

Fig. 8. A 25-s simulated wind speed profile.

## IV. Simulation Results

The type-4 wind turbine with HESS is simulated on MATLAB/Simulink. The PMSG has a rated power of 1.5 MW, $R_s = 0.006 \, \Omega$, $L_d = L_q = 0.3$ mH. The flux linkage is $\psi_m = 1.48$ V·s, the number of pole pairs is $p = 40$. The equivalent momentum of inertia is $J = 3.5 \cdot 10^4 \, kg \cdot m^2$. The feedback signals and control loops are sampled and executed, respectively, every $100 \, \mu s$. The line-line voltage at the grid side is 575 V root-mean-squared (RMS), and the DC-link voltage is maintained at

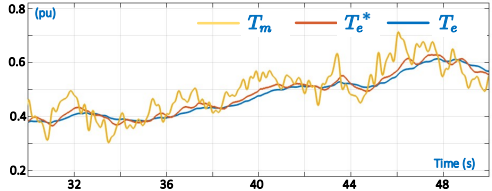
Fig. 9. Comparison between the mechanical torque, the commanded electrical torque, and the electrical torque.

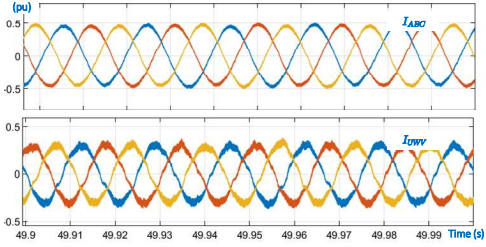
Fig. 10. The three-phase currents from the grid-side (ABC) and the machine-side (UWV).

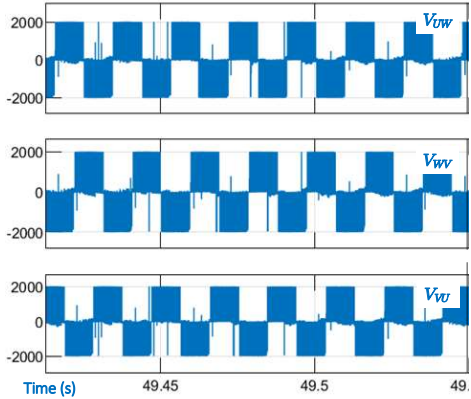
Fig. 11. The output line-line voltage in the motor side.

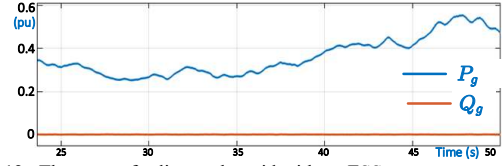
Fig. 12. The power feeding to the grid without ESS.

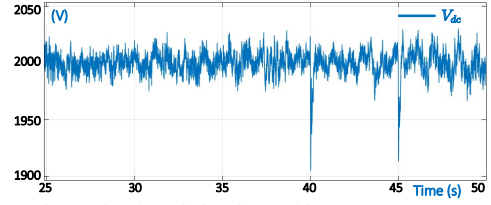
Fig. 13. The regulated DC-link voltage without HESS.

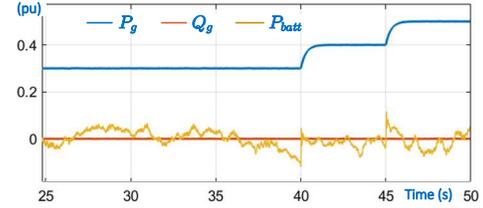
Fig. 14. The grid power and the battery power with battery.

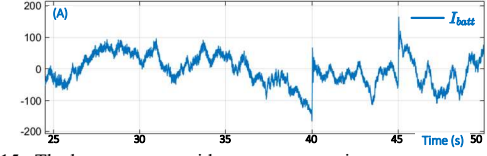
Fig. 15. The batter current without super-capacitor.

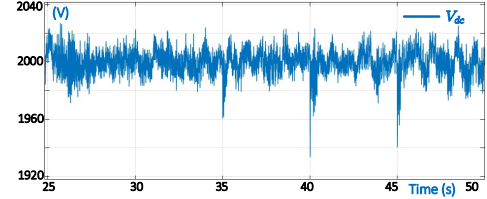
Fig. 16. The regulated DC-link voltage with battery.

2000 V. The rectifier and inverter are combined following the proposed dual-buck 9-switch converter. The battery and super-capacitor are connected to the DC-link by the proposed bidirectional dual-buck 3-switch converter. The battery has a nominal voltage of 750 V and a rated capacity of 200 kWh. The super-capacitor bank has a nominal voltage of 1250 V, a rated capacity of 100 kWh with $C_{max} \approx 1000\ F$ [19].

The wind speed includes four components: mean wind speed, wind gust, wind ramp, and wind noise. The wind profile shown in Fig. 8 is simulated with a mean value of $6\ m/s$. In Fig. 9, the mechanical torque, the commanded electrical torque, and the measured electrical torque is compared. The regulated three-phase current measured from the grid-side and the motor side is shown in Fig. 10. In Fig. 11, the output line-line voltages of the proposed 9-switch power converter on the motor- side is shown. These waveforms validate the operation performance of the proposed 9-switch power converter system.

There are three case studies that are compared as follows: 1) without HESS, 2) with only battery, and 3) with both battery and super-capacitor.

### A. Without HESS.

The power feeding to the grid in case of no HESS is shown in Fig. 12. As can be seen, that power follows the variations of the wind speed profile and commanded electrical torque. The DC-link voltage in this case is still regulated at 2000 V as shown in Fig. 13. There are some down spikes at 40 s and 45 s due to the high variations of the wind speed.

### B. With Battery.

The power feeding to the grid in case of using the battery is shown in Fig. 14 in comparison with the battery power. As can be seen, the grid power can be regulated at a desired level of 0.3, 0.4 and 0.5 (pu) in this case. The excessive and the lacking powers are absorbed or replenished by the battery, respectively. As shown in Fig. 15, the battery current is highly variable with some big spikes of more than 100 A. This may reduce the battery life. The regulated DC-link voltage is shown in Fig. 16 with smaller variations compared to those in Fig. 13

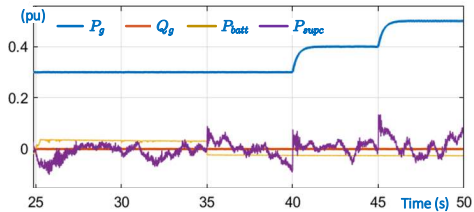
Fig. 17. The grid power, battery power and super-capacitor power with battery and super-capacitor.

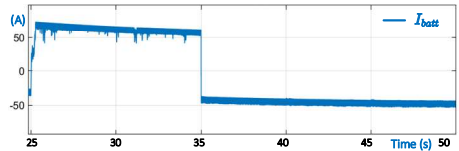
Fig. 18. The battery current with super-capacitor supporting.

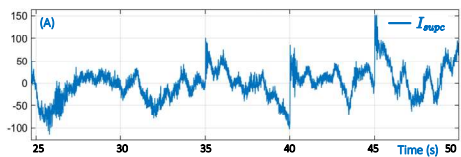
Fig. 19. The super-capacitor current.

without HESS.

*C. With Battery and Super-Capacitor (HESS)*.

Adding the super-capacitor supports the CC and CV charge and discharge mode of the battery. Therefore, the battery life can be increased. The grid power, battery power, and supercapacitor power are shown in Fig. 17. As can be seen, the battery power is regulated constantly with a nearly constant current of 50 A and -50A for charge and discharge, respectively, as shown in Fig. 18. The remaining power is taken care of by the super-capacitor. The super-capacitor current is shown in Fig. 19.

## V. CONCLUSION

The paper proposed the integrated multi-port power converter configuration for the bidirectional back-to-back system of the type-4 wind turbine with a hybrid energy storage system included battery and super-capacitor. The proposed power converter system can reduce 4 active switches and the accompanied gate-driver circuits and auxiliary power supplies compared to the traditional configuration. With the dual-buck structure embedded, the proposed three-switch leg of the power converter can withstand the short-circuited. Therefore, the proposed system is more resilient. The operation performance of the proposed power converter configuration is analyzed theoretically and validated by simulation results. The advantages of the hybrid energy storage system with battery and super-capacitor was also proved.

## ACKNOWLEDGMENT

The information, data, or work presented herein was funded in part by the New York State Energy Research Development Authority under award number 148516.